\documentclass[useAMS,usenatbib,usegraphicx]{mn2e}
\usepackage{graphicx,times,epsfig}

\newcommand{\kmsend}{\mbox{km s$^{-1}$}}

\newcommand{\hmsun}{\mbox{h$^{-1}$M$_{\sun}$ }}

\newcommand{\msun}{\mbox{M$_{\sun}$ }}
\newcommand{\msunend}{\mbox{M$_{\sun}$}}

\newcommand{\msunyr}{\mbox{M$_{\sun}$yr$^{-1}$ }}
\newcommand{\msunyrend}{\mbox{M$_{\sun}$yr$^{-1}$}}
\newcommand{\htwo}{\mbox{H$_2$}}

\newcommand{\z}{\mbox{$z$}}

\newcommand{\zsim}{\mbox{$z\sim$ }}

\newcommand{\sef}{\mbox{S$_{\rm 850}$}}

\newcommand{\sunrise}{\mbox{\sc{sunrise}}}
\newcommand{\mappings}{\mbox{\sc{mappingsiii}}}
\newcommand{\gadget}{\mbox{\sc{gadget3}}}
\newcommand{\microjy}{\mbox{$\mu$Jy}}
\title[The Formation and Evolution of SMGs]{The
  Formation of High Redshift Submillimeter Galaxies}

\author[Narayanan, Hayward, Cox, Hernquist, Jonsson, Younger \&
  Groves]{Desika\, Narayanan$^{1}$\thanks{E-mail:
    dnarayanan@cfa.harvard.edu}\thanks{CfA Fellow}, Christopher\, C.\,
  Hayward$^{1}$, Thomas\, J.\, Cox$^{1}$\thanks{W.M. Keck Postdoctoral
    Fellow}, Lars\, Hernquist$^{1}$, \and Patrik\, Jonsson$^{2}$,
  Joshua\, D.\, Younger$^{1,4}$\thanks{Hubble Fellow}, and Brent
  Groves$^{3}$\\$^{1}$Harvard-Smithsonian Center for Astrophysics, 60
  Garden St., Cambridge, Ma 02138\\$^{2}$Santa Cruz Institute for
  Particle Physics, University of California, Santa Cruz, Santa Cruz,
  Ca\\$^{3}$Sterrewacht Leiden, Neils Bohrweg 2, Leiden 2333-CA, The
  Netherlands\\$^4$Current Address: Institute for
  Advanced Studies, Einstein Drive, Princeton, NJ 08544}
\begin{document}

\date{Accepted by MNRAS}

\pagerange{\pageref{firstpage}--\pageref{lastpage}} \pubyear{2009}

\maketitle

\label{firstpage}

\begin{abstract}

We describe a model for the formation of \zsim 2 Submillimeter
Galaxies (SMGs) which simultaneously accounts for both average and
bright SMGs while providing a reasonable match to their mean observed
spectral energy distributions (SEDs). By coupling hydrodynamic
simulations of galaxy mergers with the high resolution 3D
polychromatic radiative transfer code \sunrise, we find that a mass
sequence of merger models which use observational constraints as
physical input naturally yield objects which exhibit black hole,
bulge, and \htwo \ gas masses similar to those observed in SMGs.  The
dominant drivers behind the 850 \micron \ flux are the masses of the
merging galaxies and the stellar birthcloud covering fraction.  The
most luminous (\sef$\ga$15 mJy) sources are recovered by
$\sim$10$^{13}$ \msun 1:1 major mergers with a birthcloud covering
fraction close to unity, whereas more average SMGs (\sef $\sim$5--7
mJy) may be formed in lower mass halos
($\sim$5$\times$10$^{12}$\msun).  These models demonstrate the need
for high spatial resolution hydrodynamic and radiative transfer
simulations in matching both the most luminous sources as well as the
full SEDs of SMGs. While these models suggest a natural formation
mechanism for SMGs, they do not attempt to match cosmological
statistics of galaxy populations; future efforts along this line will
help ascertain the robustness of these models.

\end{abstract}


\section{Introduction}

As a population, \zsim 2 Submillimeter Galaxies (SMGs) are the most
luminous, heavily star-forming galaxies in the Universe
\citep{bla02}. The last decade of observations have provided
significant constraints regarding their physical properties. Selected
for their prodigious long wavelength emission ($\ga$5 mJy at 850
\micron), SMGs are thought to be starburst dominated
\citep{cha04,ale08,you08b}, and have massive
$\sim$10$^{10}$--10$^{11}$\msun \ \htwo \ gas reservoirs
\citep{tac06,tac08,gre05}. If stars form according to a Kroupa initial
mass function (IMF), as is supported by observations \citep{tac08},
then the infrared luminosities from SMGs translates to tremendous
inferred $\sim$700--1500 \msunyr star formation rates
\citep{kov06,pop08} driven by major mergers \citep{tac08}. SMGs are
extremely massive, with typical $\sim$5$\times$10$^{11}$\msun bulges
in place \citep{bor05,swi04}, and are a highly clustered population of
galaxies, residing in $\sim$10$^{12}$--10$^{13}$\msun dark matter halos
\citep{bla04,swi08}.

Understanding SMGs from a theoretical standpoint has proven a major
challenge for models. Recently, by assuming a flat stellar initial
mass function (IMF) for starbursting galaxies (such that d$n$/d(ln
$\ m)=m^{-x}$ with $x=$0 as compared to 1.35 for Salpeter's
IMF)\footnote{Though the issue of the IMF in \zsim 2 SMGs is
  controversial \citep{bau05,swi08,tac08,dav08,van08}, we do not
  attempt to investigate the role of the IMF in reproducing
  cosmological number counts in this paper; this will be investigated
  in due course.}, semi-analytic models (SAMs) of galaxy formation
coupled with 2-D spectrophotometric simulations \citep{sil98} have
successfully reproduced the observed SMG number counts at \zsim 2.
Though these models have struggled to match some details of SMGs
\citep[e.g. the $K$-band through mid-IR SEDs;][]{swi08}, it is clear
from these models that, in principle, SMGs broadly fit into our
current understanding of high-\z \ galaxy formation.

However, while the SAMs provide insight into the cosmological
properties of SMGs, their usage of simplified analytic prescriptions
for star formation and AGN activity mean that detailed information on
the formation, evolution, and physical properties of individual SMGs
is lacking. In this arena, high resolution hydrodynamic simulations
are better suited, yet thus far have struggled. While recent
theoretical work utilizing gas rich galaxy merger simulations has
shown good success in reproducing the observed characteristics of a
number of galaxy populations -- e.g. quasars \citep[][ and references
  therein]{hop05b,hop06}, red galaxies \citep{spr05c,hop06b,hop08b},
elliptical galaxies \citep{cox06b,hop08a,hop08c,hop09b,hop09c}, and
warm Ultraluminous Infrared Galaxies \citep{you09} -- a
detailed model for SMG formation and evolution using simulations has
remained elusive. For example, attempts at forming SMGs by combining
hydrodynamic and radiative transfer calculations of gas rich mergers
have shown a peak flux of only $\sim$4--5 mJy in the submillimeter for
the most massive ($M_{\rm DM} \approx$ 2$\times$10$^{13}$) and
luminous sources \citep{chak08}. This is only marginally detectable in
current deep, wide cosmological surveys \citep{cop06}, and has dark
matter and stellar masses too large to be representative of the
average SMG population \citep{ale08,swi08}.
 % While one
%  could imagine that
%trivially increasing the
%galaxy mass would increase the simulated submillimeter flux, this
%would quickly violate the halo mass constraints inferred by clustering
%measurements of SMGs \citep{bla04}.

This Letter is the first in a series of papers attempting to
understand the detailed properties of SMGs using simulations.  Here,
we present the first model for the formation of SMGs that reproduces
both average and bright SMGs while plausibly matching their mean
observed SEDs, and stellar bulge, \htwo, and black hole masses.  We
show that high resolution hydrodynamic and radiative transfer
simulations are necessary to capture both the requisite starbursts as
well as stellar bulge growth and dust obscuration to simultaneously
power both the most luminous SMGs and match the observed SEDs.
Throughout this paper, we assume $\Omega_\Lambda=0.7$, $\Omega_M=0.3$,
and $h=0.7$.

\section{Numerical Methods}
\label{section:methods}

We follow the hydrodynamic evolution of galaxy mergers of varying mass
and mass ratios using the $N$-body plus smoothed particle
hydrodynamics (SPH) code \gadget \  \citep{spr05b}. The synthetic
photometric properties of the simulations are then calculated using
the 3D adaptive mesh Monte Carlo polychromatic radiative transfer
code, \sunrise \ \citep{jon06a}. Our goal throughout is to utilize
physical parameter input to the models directly constrained by
observations of SMGs where possible, and from local ULIRGs or the
Milky Way Galaxy when data from SMGs may not be available.

\subsection{Hydrodynamics}
 \gadget \ utilizes an entropy-conserving formalism for SPH
 \citep{spr02} which include prescriptions for radiative cooling of
 the gas \citep{kat96, dav99}, and a sub-grid formalism for a
 multi-phase interstellar medium (ISM) in which cold clouds are
 embedded in a hot, pressure confining phase \citep{spr03a}. The cold
 clouds are allowed to grow through radiative cooling of the hot phase
 gas, and similarly, supernovae may evaporate the cold gas into the
 hotter phase. Numerically, supernovae pressurization of the ISM is
 handled through an effective equation of state (EOS). Here, we employ
 a stiff EOS with $q_{\rm EOS}=1$ \citep[see Fig 4. of
 ][]{spr05a} which allows for stable disk evolution.  Star formation
 in the simulations proceeds following a generalized Kennicutt-Schmidt
 (KS) relation (e.g. SFR $\propto \rho^{1.5}$) with normalization set
 to match the local surface density relation. This assumption is
 bolstered by observations of SMGs which hint at a KS index near the
 locally observed value, 1.5 \citep{bou07}.

Black holes are included in the simulations as sink particles that
accrete surrounding material following an Eddington limited
Bondi-Hoyle-Lyttleton paramaterization \citep[e.g.][]{bon44}. A
subresolution model for feedback is incorporated as an isotropic
thermal coupling between the active galactic nucleus (AGN) and the
surrounding ISM.  The efficiency of this coupling is tuned to match
the local M-$\sigma$ relation \citep{dim05}.

The galaxies are initialized with a \citet{her90} dark matter profile,
and virial properties scaled to be appropriate for redshift \zsim 3
\citep{rob06b}. The galaxies are initialized with circular velocities
$V_{\rm circ}=320--500$ \kmsend, motivated by the circular velocities
for SMGs measured by \citet{tac08}. This results in halo masses of
$M_{\rm DM} \sim$10$^{12}--10^{13}$ \msunend, consistent with SMG
clustering measurements \citep{bla04}.  The galaxies are bulgeless,
and the initial gas fraction is 80\%. This results in galaxies with
gas fractions $\la$ 40\% at final coalescence (owing to gas
consumption by star formation), comparable to estimates of observed
SMGs \citep{bou07}. Our gas particle masses were $M=1.4\times$10$^6$
\hmsun, and softening lengths 100 $h^{-1}$pc. In this paper, we consider 12
merger models varying mass, merger orbit and mass ratio. We consider
1:1 mergers in $\sim$10$^{12}$, 5$\times$10$^{12}$ and 10$^{13}$\msun
halos, and 1:3 and 1:12 mergers with a $M_{\rm DM} \approx $10$^{13}$
\msun \ primary galaxy.
%Because the results are generally similar across varying orbits, we
%present the results in terms of a particular fiducial orbit with
%arbitrary orbital angle $\theta_1,\phi_1$=(30,60)$^o$,
%$\theta_2,\phi_2$=(-30,45)$^o$.  In Table~\ref{table:resulttable}, we
%summarize the merger models employed here which formed SMGs, as well
%as a number of physical properties derived from the simulations (to
%be discussed in \S~\ref{section:results}) during the phase when the
%galaxy is detectable as an SMG.

\subsection{Radiative Transfer}
\sunrise \ considers the propagation of radiation between UV and
millimeter wavelengths through a dusty medium (see \citet{jon06a} for
greater detail, as well as \citet{jon09} for
important updates, some of which are summarized here). \sunrise \ 
tracks model photon 'packets' originated by sources -- AGN, stellar
clusters, and the ISM itself -- as they traverse the dusty ISM using a
Monte Carlo methodology.  As in galaxies in nature, the flux emergent
from the model galaxy/galaxies is determined by the photons that are
emitted from the stars or AGN in the camera direction and escape
without interacting and also the photons that scatter or are
re-emitted by dust into the camera direction. Here, we utilized 8
different cameras distributed uniformly around the model
galaxy/galaxies. \sunrise \  treats dust self-absorption and
re-emission in calculating the dust temperature.

New for the simulations here, the AGN emits an empirical template SED
derived from observations of unobscured quasars \citep{hop07},
assuming a radiative efficiency of 10\%.  The normalization of this
input spectrum is set by the total bolometric luminosity of the
central black hole. The spectrum emitted from stellar clusters is
calculated utilizing {\it Starburst 99} \citep{lei99}.  We assume a
Kroupa IMF for the stars, which is in good agreement with the
estimates for \zsim 2 galaxies \citep{dav08,tac08,van08}.  For young
stellar clusters, which are still located in their nascent birthclouds
of molecular gas, the input spectrum includes the effects of the HII
regions and photodissociation regions (PDRs) surrounding the clusters
\citep{gro08}.  The models describe the evolution of the HII regions
and PDRs analytically and use the photoionization code \mappings \ \citep{gro04} to calculate the SED that emerges from the
cluster after it has been reprocessed by the HII regions and PDRs.
The HII region absorbs almost all of the ionizing radiation and is
responsible for the hydrogen emission lines and hot dust emission. The
PDR absorbs a significant fraction of the UV stellar light and
reprocesses it to FIR emission via cooler dust.

\begin{figure}
\includegraphics[scale=0.35,angle=90]{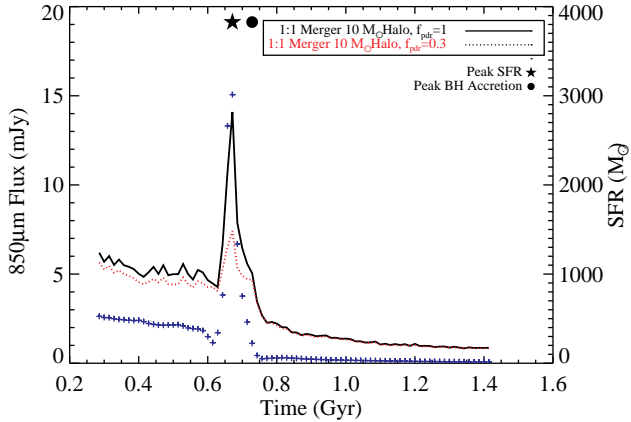}
\caption{Evolution of 850 \micron \ flux from model 1:1 major merger
  in a $\sim$10$^{13}$ \msun \ halo. Two different birthcloud covering
  fractions (f$_{\rm pdr}$) are shown (black solid line and red dashed
  line). The blue crosses represent the SFR with units on the right
  axis. The black star and dot near the top axis denote the peak of
  the starburst and black hole accretion rate, respectively. During the
  peak starburst, the opacity is dominated by stellar birthclouds, and
  the galaxy may be seen as an SMG. The most massive mergers with
  birthcloud covering fractions of unity may represent the most
  luminous observed SMGs.  \label{figure:pdrplot}}
\end{figure}

A fraction (f$_{\rm PDR}$) of clusters with ages $<$ 10 Myr are
modeled as obscured by their nascent birth clouds.  As this fraction
is increased, more UV and optical flux is reprocessed to longer
wavelengths.  The covering fraction affects the maximum sub-mm
luminosity during the peak of the starburst.  This is equivalent to a
cloud clearing time scale.  While f$_{\rm pdr} \approx 0.2$ (which
corresponds to a clearing timescale $\sim$2 Myr) may be a reasonable
approximation for field galaxies \citep{gro08}, in the case of
massive, gas rich galaxy mergers, an assumption of f$_{\rm PDR}
\approx$ 1 ($t_{\rm clear} >> $ 10 Myr) may be more
appropriate. Indeed, observations of molecular gas in local ULIRGs
have found that the nuclear regions of mergers are often best
characterized by a uniform molecular medium, rather than discrete
clouds \citep{dow98,sak99}. While this is not the same as a birthcloud
covering fraction of unity, tests done for this work have shown that
they result in similar FIR/submm fluxes\footnote{We have run
  experiments assuming f$_{\rm pdr}=0$, and that each grid cell is
  uniformly filled with dust with mass set by the ULIRG dust to gas
  ratio. In this limiting case, we find fluxes quite similar to those
  in simulations where the PDR covering fraction is unity.}. We note
that the simulations do not account for the obscuration of stellar
light by GMCs outside of the birthcloud.

%As radiation encounters a patch of dusty ISM, it may either pass
%through unhindered, scatter off of dust grains, or be absorbed by dust
%grains and reemitted at the characteristic temperature of the dust.
%{\it Sunrise} calculates the reradiated emission from dust using an
%iterative procedure to achieve radiative equilibrium. Practically,
%this means that when radiation from stellar clusters and black holes
%is absorbed by dust, the dust temperature is modified based on the
%balance between absorbed energy and reradiated emission. This process
%is repeated until the dust temperatures converge. This iteration is
%especially important when the ISM is optically thick to infrared dust
%emission, as is the case during the SMG phase of our simulations.

We utilize 10$^7$ photon packets per iteration in the radiative
equilibrium calculations, and the analysis is done for rest frame
wavelengths 0.1 to 1000 $\mu$m.  The stellar particles initialized
with the simulation are assumed to have formed at a constant rate over
$\sim$250 Myr. Metal enrichment is tracked from supernovae using an
instantaneous recycling approximation. The gas and stars initialized
with the simulation are assigned a metallicity according to a
closed-box model such that $Z=(-y$ \ ln[$f_{\rm gas}$]) where $Z$ is
the metallicity, $y$ the yield=0.02, and $f_{\rm gas}$ is the initial
gas fraction (though note the fluxes during the SMG phase of the model
galaxies is not very sensitive to these assumptions). Without
knowledge of the dust properties of \zsim 2 SMGs, we conservatively
assume a 3.1\_60 \citet{dra07} model.  We additionally assume a
constant dust to gas ratio comparable to that of ULIRGs
\citep[1/50;][]{wil08b} which tentative evidence suggests is
appropriate for SMGs \citep{kov06,gre05,tac06}. We have additionally
run models assuming a Milky Way dust to metals ratio of 0.4
\citep{dwe98}, and found similar results to within 10\%. We note that
comparisons to templates which include stochastically heated grains
\citep{dra07} suggest a potential uncertainly in the FIR SED up to a
factor of 2 \citep{jon09}.

\section{Results}
\label{section:results}

%******************************
%  Table 1: models and results
%==============================

\begin{table*}
\label{table:resulttable}
\centering
\begin{minipage}{100mm}
\caption{}
\begin{tabular}{@{}cccccc@{}}
\hline Total $M_{\rm DM}$ & Mass Ratio & $M_{\star}$ & $M_{\htwo}$ &
$M_{\rm BH}$ \footnote{Owing to long dynamical friction time scales, minor
  mergers tend to be relatively gas poor upon final coalescence, and
  show their largest burst of 850 \micron \ flux on first
  passage. While BH growth is generally self-regulated, and the final
  coalescence BH masses are robust for a large range of initial BH seed
  masses, $M_{\rm BH}$ for first passage starbursts can vary wildly
  based on initial seed masses \citep{you08a}. As such, the BH masses
  for first-passage SMGs in the simulations are relatively
  uncertain.} & Peak \sef \\
$ M_{\odot}$&&$M_{\odot}$&$M_{\odot}$&$M_{\odot}$&mJy\\
\hline
2$\times$10$^{13}$&1:1&8$\times$10$^{11}$&1.5$\times$10$^{11}$&3.3$\times$10$^{8}$&14.1\\
1.1$\times$10$^{13}$&1:3&2.8$\times$10$^{11}$&3$\times$10$^{10}$&*&6.6\\
7$\times$10$^{12}$&1:1&2$\times$10$^{11}$&1.8$\times$10$^{10}$&0.8$\times$10$^{8}$&5.6\\
9.8$\times$10$^{12}$&1:12&2.9$\times$10$^{11}$&2$\times$10$^{10}$&*&4.8\\
\hline

\end{tabular}
\end{minipage}
\end{table*}

%\begin{deluxetable}{cccccc}
%\tabletypesize{\scriptsize} \tablecaption{SMG Physical
%  Parameters\label{table:resulttable}}
%\tablewidth{0pt}
%\tablehead{
% \colhead{Total $M_{\rm DM}$} & \colhead{Mass Ratio}  &
%\colhead{$M_{\star}$} & \colhead{$M_{\htwo}$} &
%\colhead{$M_{\rm BH}$} & \colhead{Peak \sef} \\
%$ M_{\odot}$&&$M_{\odot}$&$M_{\odot}$&$M_{\odot}$&mJy\\
%}%close tablehead
%
%\startdata
%2$\times$10$^{13}$&1:1&8$\times$10$^{11}$&1.5$\times$10$^{11}$&3.3$\times$10$^{8}$&14.1\\
%1.1$\times$10$^{13}$&1:3&2.8$\times$10$^{11}$&3$\times$10$^{10}$&*&6.6\\
%7$\times$10$^{12}$&1:1&2$\times$10$^{11}$&1.8$\times$10$^{10}$&0.8$\times$10$^{8}$&5.6\\
%9.8$\times$10$^{12}$&1:12&2.9$\times$10$^{11}$&2$\times$10$^{10}$&*&4.8\\
%
%
%\tablenotetext{*}{Owing to long dynamical friction time scales, minor
%  mergers tend to be relatively gas poor upon final coalescence, and
%  show their largest burst of 850 \micron \ flux on first
%  passage. While BH growth is generally self-regulated, and the final
%  coalescence BH masses are robust for a large range of initial BH seed
%  masses, $M_{\rm BH}$ for first passage starbursts can vary wildly
%  based on initial seed masses \citep{you08a}. As such, the BH masses
%  for first-passage SMGs in the simulations are relatively
%  uncertain. }
%
%\end{deluxetable}

%==============================

\subsection{Formation of a Submillimeter Galaxy}

Galaxy mergers residing in halos with masses comparable to those
observed naturally produce submillimeter galaxies with a range of 850
\micron\ fluxes during their peak starburst phase. Mergers with a
total halo mass above $\sim$5$\times$10$^{12}$\msun will produce
galaxies above the nominal SMG detection threshold (\sef \ $\ga$ 5 mJy)
whereas significantly lower mass mergers may not. The peak observed
submillimeter flux is dependent on the total dust obscuration for a
given merger model.

\begin{figure}
\includegraphics[scale=0.40,angle=90]{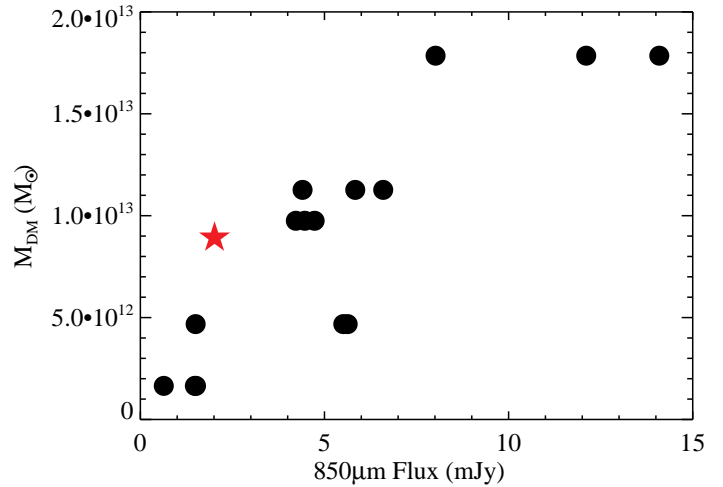}
\caption{Total halo mass versus peak \sef. The full range of observed
  850 \micron \ fluxes from SMGs may be understood from a mass
  sequence merger models, varying only total mass.  More massive
  mergers induce stronger bursts which fuel more luminous SMGs.  The
  red star shows an isolated disk.  Some merging activity may be
  necessary to induce the necessary SFRs to form SMGs. Even massive
  ($\sim$10$^{13}$\msunend) isolated disks seem unable to reach the
  nominal \sef \ $> 5$ mJy detection threshold, though may be routinely
  detectable via future-generation bolometers.\label{figure:massplot}}
\end{figure}

To see this, in Figure~\ref{figure:pdrplot}, we begin by showing the
evolution of the 850 \micron \ flux \citep[at \z \ = 2, the rough mean
  redshift for radio-selected SMGs;][]{cha03a} for a M$_{\rm DM}
\approx 10^{13}$ \msun 1:1 merger for two birthcloud covering
fractions. Additionally, we overplot the SFR in light blue
symbols. During the first passage and in-spiral phase of a galaxy
merger, the galaxies are forming stars (individually) at $\sim$100--200
\msunyrend. During this time, obscuration by the diffuse ISM dominates
and produces a modest $\sim$5 mJy of emission seen during the in-spiral
phase, regardless of the GMC covering fraction.

As the galaxies reach final coalescence (T$\sim$0.65 Gyr), tidal
torquing of the gas fuels a $\sim$2000 \msunyr nuclear starburst. At
this time, the FIR/submillimeter flux is dominated by cold dust
reprocessing of UV photons in young stellar clusters by
their birthclouds. Hence, the submillimeter flux rises rapidly with this
burst in star formation rendering the galaxy detectable as an
extremely luminous SMG for a relatively short ($<$50 Myr)
lifetime. The peak flux of the SMG produced in the final burst is
dependent on the birthcloud covering fraction.  The merger model in
Figure~\ref{figure:pdrplot} with a covering fraction of 30\% reaches a
peak flux of $\sim$7--8 mJy, whereas the model with covering fraction
of unity produces an SMG comparable to the most luminous observed
sources. The obscuration of young stars is vital to creating the most
luminous SMGs as these stars dominate the bolometric luminosity of the
galaxy during coalescence.

Because the peak flux received is primarily determined by the strength
of the starburst (for a given birthcloud covering fraction), the
submillimeter flux is dependent on the mass of the merger. Less
massive galaxies will induce smaller tidal torques on the gas, fueling
lower SFRs; in these cases, the lightcurve shown for the massive
merger in Figure~\ref{figure:pdrplot} will retain its shape, though
simply lower in normalization. In Figure~\ref{figure:massplot}, we
indicate this more explicitly by showing the predicted 850 \micron
\ flux against the galaxy (halo) mass of all 12 merger models varying
primary halo mass, mass ratio (1:1, 1:3 and 1:12) and orbit. Again,
the 850 \micron \ flux is modeled at \z \ = 2. A single isolated disk is
shown by the red star.  Each is assumed to have a constant birthcloud
covering fraction of unity.

The 1:1 major mergers in the most massive ($M_{\rm DM}\approx2\times
10^{13}$\msun) halos produce SMGs comparable to the most luminous
observed \citep{cop06,pop06}. Galaxies $\sim$2--4 times smaller may be
representative of more average (\sef\ $\sim 5$ mJy) SMGs. Galaxies well
below this mass limit are unlikely to ever produce SMGs based on the
current fiducial criteria \sef \ $\ga$5 mJy, though may be detectable
with sensitive new bolometer arrays.  As is shown by the isolated
galaxy (red star, Figure~\ref{figure:massplot}), some merging seems to
be necessary to fuel the starbursts that power SMGs. Isolated galaxies
residing in even the most massive halos in our simulations ($M_{\rm
  DM} \approx10^{13}$ \msun) never appear to get above $\sim$2--3 mJy
at 850 \micron.

\subsection{Physical Properties and SEDs of SMGs}

The SMGs produced by the galaxy mergers presented here have black
hole, stellar, \htwo, and dark matter masses (the latter by
construction) comparable to those observed. In
Table~\ref{table:resulttable}, we show these values for a typical
orbit ($\theta_1,\phi_1=(30,60)^o$, $\theta_2,\phi_2=(-30,45)^o$)
for the simulations which produced detectable SMGs.  As shown in Cox
et al. (2009, submitted) and \citet{nar08a}, binary galaxy mergers
produce $\sim$90\% of their final stellar mass during the
pre-coalescence phase. As such, the bulk of the $\sim$10$^{11}$\msun
bulges are already in place during when the galaxy is visible as an
SMG. Indeed, this is clear from a visual integration of the SFR in
Figure~\ref{figure:pdrplot}; an SFR of a few$\times$10$^{2}$ \msunyr
for $\sim$5$\times$10$^{8}$ years results in a few$\times$10$^{11}$
\msun bulge.

The black hole growth follows a different story: the bulk of the
growth occurs concomitant to the SMG phase/final coalescence.  While
the final black hole masses in the most massive/luminous systems may
be similar to those of \zsim 2 quasars ($\sim$10$^9$ \msun), during
the SMG phase they can be a factor of $\sim3--10$ lower than these
final values. This gives them masses of order $\sim$10$^8$\msun,
comparable to observed values \citep[][and references therein]{ale08}.
Finally, as was noted in \S~\ref{section:methods}, by starting with
progenitor galaxies with gas fractions $f_g=0.8$, the final gas
fraction of during the SMG phase is $\la$40\%, comparable to estimates
of observed SMGs \citep{bou07}. While \gadget \ does not track the
evolution of molecular gas, we derive the molecular gas fraction in
post-processing based on the ambient pressure on the cold,
star-forming ISM \citep[see][]{bli06}. From this, we find total \htwo
\ gas masses between $10^{10}--10^{11}$ \msun
(Table~\ref{table:resulttable}). This is comparable to measurements by
\citet{gre05} and \citet{tac06}. Moreover, complementary radiative
transfer simulations by \citet{nar09b} utilizing this methodology for
assigning the \htwo \ gas fraction have shown good correspondence
between the simulated CO properties (e.g. excitations and line widths)
of these model SMGs, and those in nature.

\begin{figure}
\includegraphics[scale=0.35,angle=90]{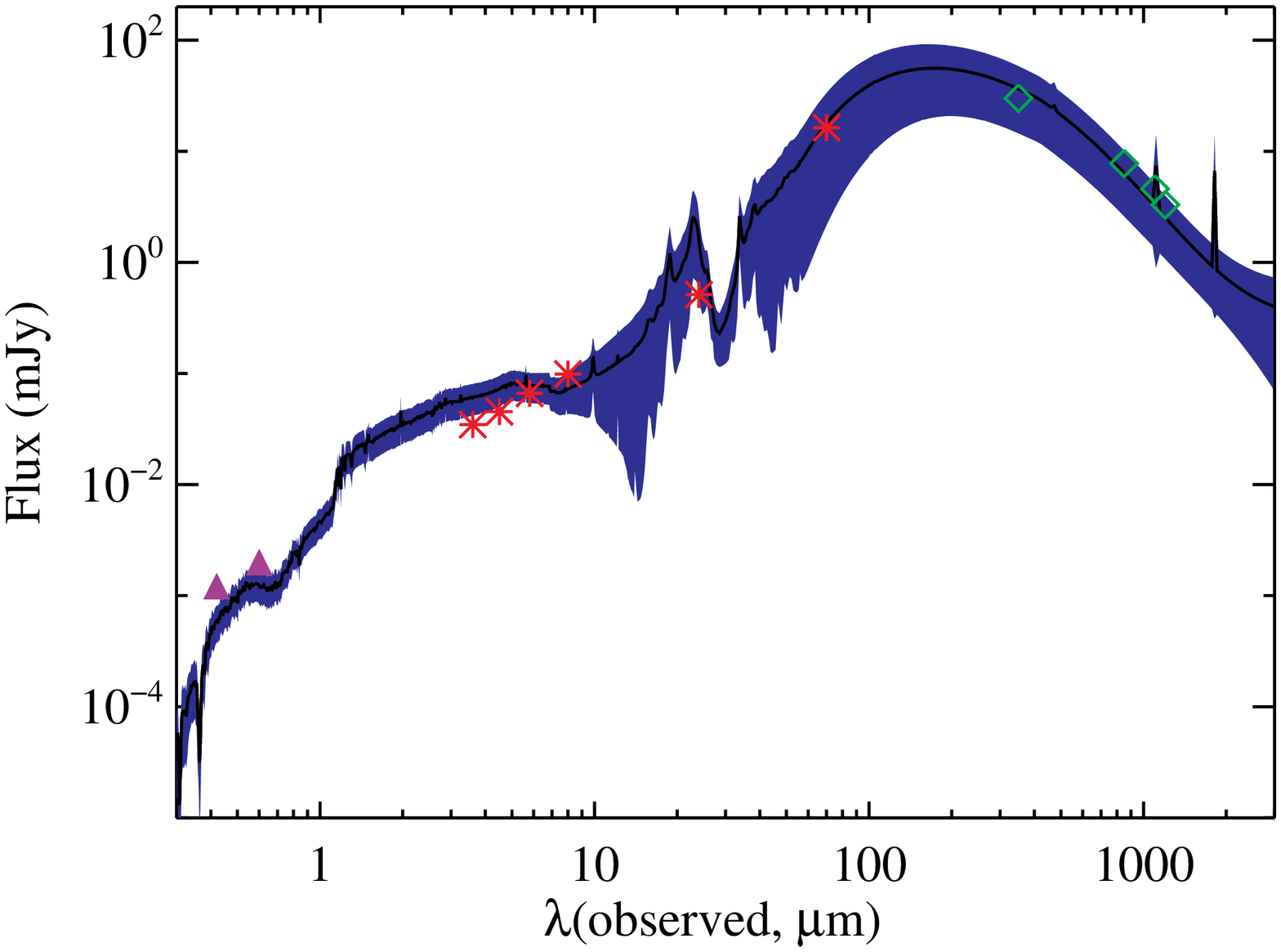}
\caption{Mean SED of all snapshots from merger simulations satisfying
  the \sef$>$5 mJy selection criterion for SMGs (solid line). The
  dispersion amongst the snapshots is shaded in blue (the dispersion
  amongst viewing angles, however, is not shown).  The simulated SED
  has been redshifted to \z \ = 2. Observed data points from
  \citet{cha05}, Hainline et al. (2009, submitted) and \citet{kov06}
  are overlaid (purple triangles, red crosses and green diamonds,
  respectively). These data all have spectroscopic redshifts.  The
  points represent the mean observed value from these surveys, though
  galaxies with $z<$1 and upper limits have been discarded. The model
  SED from the fiducial merger provides a good match to the observed
  photometric data from the $B$-band ~ to 1 mm, making this the first
  model SMG to reproduce the full optical through mm-wave
  SED.\label{figure:sed}}
\end{figure}

Galaxy mergers in a mass range of M$_{\rm DM} \approx $5
$\times$10$^{12}$\msun--10$^{13}$\msun naturally produce SMGs with
optical through mm-wave SEDs comparable to those observed. In
Figure~\ref{figure:sed}, we show the model SED for all snapshots in
the merger simulations studied here which produce an SMG with \sef $>$
5 mJy, redshifted to \z \ = 2.  The blue shaded region shows the
1$\sigma$ dispersion amongst all snapshots which satisfy this fiducial
selection criterion while the black solid line shows the mean amongst
snapshots. The mean observed data points from the surveys of
\citet[][purple triangles]{cha05}, Hainline et al. (2009, red
crosses), and \citet[][green diamonds]{kov06} are overlaid. Only
observational data with spectroscopic redshifts are utilized.

The SED encodes information regarding the young stars (at \zsim 2,
observed-frame $K$-band), stellar mass (observed mid-IR), and cold
dust obscuration levels (observed submillimeter). The SFR and stellar
bulge growth are explicitly tracked in the progenitor galaxies and SMG
by the high resolution SPH simulations while the dust obscuration is
modeled globally by diffuse dust, and on local scales by the subgrid
implementation of the \mappings \ photoionization
calculations. Because the SFRs in the simulated galaxies are
comparable to those observed, and bulges of order
$\sim$1--8$\times$10$^{11}$ \msun are in place during the SMG phase
(see Table~\ref{table:resulttable}), the observed $K$-band and mid-IR
fluxes in the model SED match the observed data points
closely. Sufficient cold dust obscuration provided by the diffuse dust
and birthclouds reprocesses the rest-frame UV emission into the cold
dust tail providing a good match to the mean observed 350, 850, and
1100 \micron \ data points. The cases with lower birthcloud covering
fraction (f$_{\rm pdr}=0.3$) tend to overestimate the rest-frame UV
and optical flux by a factor of $\sim$2.

While our simulations do not include radio emission, we can estimate
the radio properties of our simulations by assuming that the
far-IR/radio correlation does not evolve strongly with redshift
\citep[e.g. ][ Younger et al. 2009, in press]{kov06,you09b,saj08}. We
estimate the FIR flux via the standard FIR = 1.26 $\times
10^{-14}(2.58 S_{\rm 60 }$+$S_{\rm 100})$ W m$^{-2}$, where $S_{\rm 60
}$ and $S_{\rm 100}$ are the observed-frame fluxes at 60 and 100
\micron, in Jy), and then relate the FIR flux to the radio (1.4 GHz)
flux through the FIR-radio correlation \citep[e.g. Equation 5 from
][]{kov06}. We assume a range of $q$-parameters, from 2.14
\citep[consistent with observations of SMGs; ][]{kov06}, to the
locally-observed value for IRAS BGS galaxies, $q=2.35$
\citep{san96}. We include bandwidth compression and a $K$-correction
\citep[assuming an SED slope of $\alpha = 0.7$, typical for
  synchrotron power laws; ][]{con92} when redshifting the radio
flux. We note that this method neglects any potential radio
contribution from the AGN. When using the observed FIR-radio
correlation for SMGs, we recover inferred 1.4 GHz flux densities (at
\z \ = 2.0) of 20--700 $\mu$Jy. The average value amongst our
simulation sample for a $q$=2.14(2.35) is $S_{\rm 1.4} \approx 125{\rm
  (}95{\rm )}$ \microjy, similar to values described in the
compilation by \citet{cha03c}. The 850/$S_{\rm 1.4}$ flux density
ratios from our simulated SMGs is additionally in good agreement with
measurements from recent surveys \citep[e.g. ][]{pop06}. For example,
we find the mean 850/$S_{\rm 1.4}$ ratio for \sef\ $> 5 $mJy SMGs to
be $\sim$60 (with a 1$\sigma$ dispersion of 40) when utilizing
$q$=2.14. The mean(dispersion) increases to 100(65) when utilizing the
local $q$.

\begin{figure}

\includegraphics[scale=0.35,angle=90]{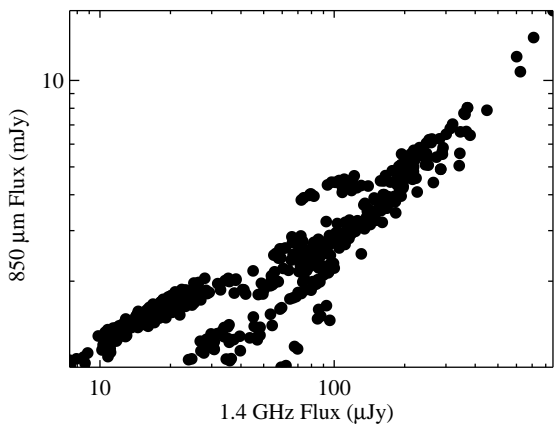}
\caption{\sef \ versus 1.4 GHz inferred flux density at \z=2.0 for all
  models in Table~\ref{table:resulttable}. Given current flux limits, the SMGs
  simulated   here    will   have   a   high    incidence   of   radio
  counterparts.\label{figure:submm_radio}}

\end{figure}

The SMGs formed in these simulations would generally have a high
detection fraction of radio counterparts. In
Figure~\ref{figure:submm_radio}, we plot the inferred 1.4 GHz flux
density (derived utilizing $q=2.14$) versus \sef \ at \z=2.0 for all
model galaxies in our simulations. We extend the nominal detection
threshold to \sef $> 1 $mJy to serve as a prediction. SMGs (\sef $> 5$
mJy) at \z=2.0 will typically be detectable down to $\sim$100 \microjy
\ at 1.4 GHz. While it is difficult to make quantitative comparisons
to observed detection fractions, owing to the fact that these
simulations are not cosmological (and \sef \ is roughly constant with
increasing redshift while $S_{\rm 1.4}$ decreases rapidly with
increasing \z), it is still feasible to make qualitative comparisons
to observations. Observed SMGs have a high incidence of radio
counterparts. Large radio ID fractions were reported by \citet{ivi02},
who found that nearly all \sef \ $>$ 10 mJy SMGs were detected above
$\sim$35 \microjy \ at 1.4 GHz, and approximately half the SMGs above
\sef \ $> 5 $ mJy were radio-detected. More recently, \citet{wag09}
find that some $\sim$80\% of the SMGs in their sample contain radio
counterparts above $\sim$25\microjy.  More generally, the dynamic
range probed by the submillimeter flux densities and radio flux
densities in Figure~\ref{figure:submm_radio} is comparable to that
seen in the \sef-$S_{\rm 1.4}$ relation compiled by \citet{cha03c}.

\begin{figure*}
\includegraphics{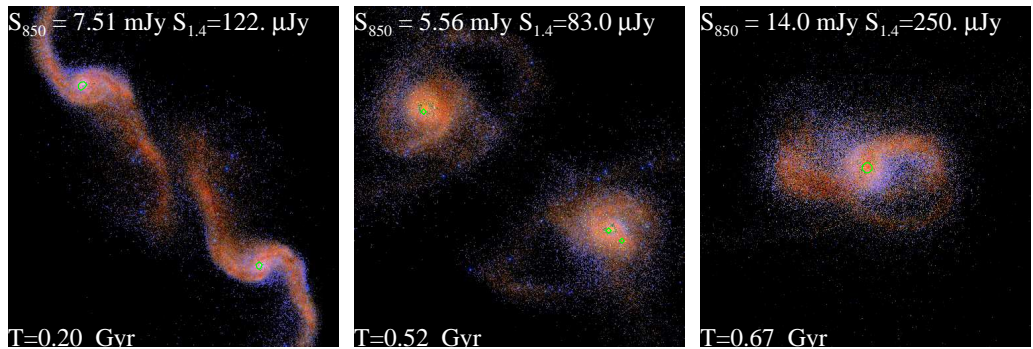}\
\vspace{1.5cm}
\caption{Optical (Sloan $\z,r,u$) colors mapped onto R,G,B for the
  most massive/luminous SMG in our simulation sample.  The thick green
  contours toward the galactic nuclei of each image enclose the region
  where the radio emission is above $S_{\rm 1.4 GHz} > 0.1 $
  \microjy. Note, the contours are only at a single level. During the
  inspiral phase of this massive ($M_{\rm DM} \approx 2 \times
  10^{13}$ \msunend) merger, the galaxy may be viewed as having
  multiple radio counterparts. \label{figure:morph}}
\end{figure*}

Many of the model SMGs in these simulations have multiple radio
counterparts. In Figure~\ref{figure:morph}, we show the optical (SDSS
$z,r,u$) morphologies of our most luminous simulated SMG through its
inspiral with a contour denoting the location of the bulk of the radio
emission (contours enclose regions where $S_{\rm 1.4} > 0.1$
\microjy). As the galaxies in Figure~\ref{figure:morph} spiral in
toward coalescence, they may be visible as multiple radio
sources. There is a trend for the more massive/luminous SMGs to have
multiple counterparts for a larger fraction of their lives. This owes
to the fact that the most massive mergers (e.g. $M_{\rm DM} > 10^{13}$
\msunend) can be SMGs during their inspiral phase
(e.g. Figure~\ref{figure:pdrplot}), whereas less massive galaxies
(e.g. $M_{\rm DM} \approx 5 \times 10^{12}$ \msunend) generally only
reach \sef \ $> 5 $ mJy when the progenitor galaxies have coalesced. To
identify individual radio sources, we search for the peaks in the radio
emission (as inferred from the FIR-radio correlation) in maps which
view the galaxies at an arbitrary angle. We then assign the peaks as
originating in separate galaxies when they originate at least 5 kpc
apart. Utilizing this prescription, we find that the least
massive/luminous SMGs in our simulations never have multiple radio
counterparts. However, the SMGs formed in $M_{\rm DM} \approx 1(2)
\times 10^{13}$ \msun halos have multiple radio counterparts
$\sim$50(75)\% of the time that they would be submillimeter-identified
(\sef $> 5$ mJy).

\section{Discussion and Conclusion}

The models presented here have shown that merger-driven SMG formation
simulations which utilize observational parameters as input produce
both average (\sef$\sim$5 mJy) and bright (\sef $\ga$ 15 mJy) SMGs
while reproducing observed physical characteristics, as well as the
full optical-mm wave SED.  This match between simulations and
observations is only possible because of the high resolution approach
employed for both the hydrodynamic and subgrid physics included in the
radiative transfer simulations.  It is worth briefly comparing these
results to previous efforts in modeling SMG formation to place our
models in context.

Models attempting to form SMGs by \citet{chak08} utilized binary
galaxy mergers similar to these. The peak submillimeter flux from
mergers comparable to the most massive observed ($M_{\rm DM}
\approx$2$\times$10$^{13}$\msunend) in this work showed $\sim$4-5 mJy
at 850 \micron; this is compared to $\sim$15 mJy from comparable mass
simulations at an identical merger orbit presented here.  While the
hydrodynamic evolution of the galaxy mergers is similar in both
studies, the radiative transfer differs significantly. In particular,
here, we employ a subgrid model for including the obscuration by the
nascent birthclouds of young stellar clusters.  As
Figure~\ref{figure:pdrplot} shows, this obscuration by cold dust
(which was not included in the \citet{chak08} simulations) is vital
during the final burst to reprocess UV photons from young stars into
FIR/submillimeter light.\footnote{\citet{chak08} incorporate a
  simplified, analytic model for the obscuring dust surrounding young
  stars.  However, in contrast to our results, they find surprisingly
  little impact on the emergent IR SED, and even a moderate increase
  the rest-frame optical and near-UV luminosity.}

The predicted SEDs from SMGs formed by semi-analytic efforts show
underluminous $K$-band and mid-IR fluxes by up to an order of
magnitude. This may owe to SFRs and stellar masses in these SMGs which
are roughly an order of magnitude lower than those inferred by
observations \citep{bau05,bau07}.  While it is difficult to directly
compare semi-analytic prescriptions with full $N$-body/SPH
simulations, it is clear that varying treatments in the gas physics
result in discrepant star formation histories, and consequent bulge
masses and simulated SEDs. The plausible correspondence between the
physical parameters and model SED from the models presented here and
those observed owes to the detailed tracking of the starburst, star
formation history, dust geometry and density distribution by these
simulations. This suggests high resolution simulations are necessary
to model the physical evolution of SMGs in detail.

That said, while the simulations presented in this work appear to
provide a reasonable match to many observed parameters from SMGs, it
is important to note that they, unlike the SAMs, {\it are not}
cosmological in nature. For a galaxy formation model to be generic, it
must concomitantly match the detailed physical and observable
properties of a given population, as well as cosmological
statistics. In this sense, the simulations presented here are
complementary to the SAMs of SMG formation. The former reproduce
detailed observables of SMGs in isolation via an {\it ab initio} model
for SMG formation, though provide no information regarding
cosmological statistics. The latter has some difficulty reproducing
e.g. the observed SED, though accurately matches the observed number
counts \citep[as well as present day $B, K$ and 60$\micron$-band
  luminosity functions;][]{bau05,swi08}. The ideal scenario would be a
coupling of the methods by convolving the lightcurves from high
resolution hydrodynamic simulations of individual mergers to
cosmological galaxy merger rates \citep[e.g.][]{hop09} to investigate
the statistical properties of SMGs. These efforts are currently are
underway.

\section*{Acknowledgments} The authors thank Sukanya Chakrabarti, 
Laura Hainline, Dusan Keres and Mark Swinbank for helpful comments. We
thank Laura Hainline for kindly providing Spitzer photometry of SMGs
in advance of publication. The simulations in this paper were run on
the Odyssey cluster supported by the Harvard FAS Research Computing
Group. This project was funded in part by a grant from the W.M. Keck
Foundation (TC), and by a Graduate Research Fellowship from the
National Science Foundation (CH). JDY acknowledges support provided by
NASA through Hubble Fellowship grant HF-51266.01 awarded by the Space
Telescope Science Institute, which is operated by the Association of
Universities for Research in Astronomy, Inc., for NASA, under contract
NAS 5-26555. PJ was supported by programs HST-AR-10678 and 10958,
provided by NASA through a grant from the Space Telescope Science
Institute, which is operated by the Association of Universities for
Research in Astronomy, Incorporated, under NASA contract NAS5-26555,
and by Spitzer Theory Grant 30183 from the Jet Propulsion Laboratory.

\bibliographystyle{apj}
\bibliography{/Users/dnarayanan/paper/refs}
\end{document}